\documentclass[AMA,Times1COL]{WileyNJDv5} 

\articletype{Article Type}%

\received{Date Month Year}
\revised{Date Month Year}
\accepted{Date Month Year}
\journal{Journal}
\volume{00}
\copyyear{2023}
\startpage{1}

\begin{document}

\title{A Risk Manager for Intrusion Tolerant Systems: Enhancing HAL 9000 with New Scoring and Data Sources}

\author[1,2]{Tadeu Freitas}

\author[1]{Carlos Novo}

\author[1]{Inês Dutra}

\author[1]{João Soares}

\author[1,2]{Manuel E. Correia}

\author[3]{Benham Shariati}

\author[4]{Rolando Martins}

\authormark{FREITAS \textsc{et al.}}
\titlemark{A Risk Manager for Intrusion Tolerant Systems: Enhancing HAL 9000 with New Scoring and Data Sources}

\address[1]{\orgdiv{Department of Computer Science}, \orgname{Faculty of Sciences of University of Porto}, \orgaddress{\city{Porto}, \country{Portugal}}}

\address[2]{\orgdiv{Centre Advanced Computing Systems}, \orgname{Institute for Systems and Computer Engineering, Technology and Science}, \orgaddress{\city{Porto}, \country{Portugal}}}

\address[3]{\orgdiv{UMBC}, \orgname{University of Maryland, Baltimore County},\orgaddress{\city{Baltimore}, \country{USA}}}

\address[4]{\orgdiv{SafeHelm, lda}, \orgaddress{\city{Porto}, \country{Portugal}}}

\corres{Corresponding author Tadeu Freitas. \email{tadeufreitas@fc.up.pt}}



\abstract[Abstract]{Intrusion Tolerant Systems (ITSs) have become increasingly critical due to the rise of multi-domain adversaries exploiting diverse attack surfaces.
ITS architectures aim to tolerate intrusions, ensuring system compromise is prevented or mitigated even with adversary presence.

Existing ITS solutions often employ Risk Managers leveraging public security intelligence to adjust system defenses dynamically against emerging threats. 
However, these approaches rely heavily on databases like NVD and ExploitDB, which require manual analysis for newly discovered vulnerabilities. 
This dependency limits the system's responsiveness to rapidly evolving threats.

HAL 9000, an ITS Risk Manager introduced in our prior work, addressed these challenges through machine learning. 
By analyzing descriptions of known vulnerabilities, HAL 9000 predicts and assesses new vulnerabilities automatically.
To calculate the risk of a system, it also incorporates the Exploitability Probability Scoring system to estimate the likelihood of exploitation within 30 days, enhancing proactive defense capabilities.

Despite its success, HAL 9000's reliance on NVD and ExploitDB knowledge is a limitation, considering the availability of other sources of information. 
This extended work introduces a custom-built scraper that continuously mines diverse threat sources, including security advisories, research forums, and real-time exploit proofs-of-concept.
This significantly expands HAL 9000’s intelligence base, enabling earlier detection and assessment of unverified vulnerabilities.

Our evaluation demonstrates that integrating scraper-derived intelligence with HAL 9000’s risk management framework substantially improves its ability to address emerging threats. 
This paper details the scraper's integration into the architecture, its role in providing additional information on new threats, and the effects on HAL 9000's management.
Our experiments show that this integration provides more secure configurations.}

\keywords{machine learning, operating system diversity, risk assessment, intrusion tolerant systems, CVE, exploits, vulnerabilities}


\maketitle

\renewcommand\thefootnote{}
\footnotetext{\textbf{Abbreviations:} ITS, Intrusion Tolerant System; CVE, Common Vulnerabilities and Exposures; CVSS, Common Vulnerability Scoring System; OSs, Operating Systems; CWE, Common Weakness Enumeration;  CPE, Common Product Enumerator; ML, Machine Learning; OSINT, Open-source Intelligence; SDN, Software Defined Networks; VM, Virtual Machine; NLP, Natural Language Processing; AI, Artificial Intelligence; SVM, Support Vector Machines; CNNs, Convolutional Neural Network; LSTM, Long Short Term Memory; XGBoost, eXtreme Gradient Boosting;  EPSS, Exploit Prediction Scoring System; API, Application Programming Interface; RMSE, Root Mean Square Deviation; Bow, bag of words; emb, sentence embeddings; NVD, National Vulnerability Database.}

\renewcommand\thefootnote{\fnsymbol{footnote}}
\setcounter{footnote}{1}

\section{Introduction}\label{sec:introduction}

Intrusion Tolerant Systems (ITSs) are used evermore to counteract and mitigate the effects of successful security breaches done to systems exposed to public networks.
The objective of ITSs is, in the event of a successful breach, to limit the adversary's capability to compromise the system, to alter its designed function, and to reduce any leakage of information, in other words, to become resilient against intrusions.
A resilient system is designed to maintain correct execution even in the presence of nodes compromised by malicious adversaries.
This is achieved by implementing methods and techniques that constrain the adversary's capabilities, ensuring the system's functionality despite breaches of the defensive perimeter.
However, the system operates with reduced performance during this period, as it actively mitigates the impact of the malicious nodes until they are fully neutralized or removed.

Most existing infrastructures and solutions rely on traditional defensive systems for their ease of deployment~\cite{techdirect_zero_trust}.
However, these systems primarily focus on securing the perimeter—the boundary between external threats and internal components or software.
As a result, they are increasingly regarded as inadequate.
Once adversaries breach this perimeter, they gain unrestricted control over the system, as no additional countermeasures are in place to mitigate or eliminate the intrusion.
The recent surge in cyber-attacks~\cite{statista_cybercrime_cost}, coupled with their increasing complexity~\cite{a1,a2,a3,a4}, where adversaries employ diversity and sophistication to achieve their objectives, has driven stakeholders to seek more effective solutions to address these threats. 
This growing demand has allowed ITS to emerge as a viable approach.

Modern ITSs are designed to address the challenges associated with the increasing scale and complexity of cyber threats. 
These systems leverage advancements from a diverse range of research domains, including, but not limited to, cybersecurity, distributed systems, and Artificial Intelligence (AI), to provide robust and adaptive solutions.
At their core, ITSs rely on replication to ensure both \textit{liveness} and \textit{security}. 
This means they can maintain correct service execution, even under compromised conditions, while safeguarding data integrity and confidentiality. 
Although this may reduce performance during adversarial activity, the service remains operational and secure.
The interdisciplinary integration of these research fields significantly enhances the resilience and robustness of ITS, enabling them to effectively mitigate the impact of malicious adversaries and adapt to evolving threats.
Current solutions now integrate distinct methods to thwart the adversaries from achieving the objective, which include, but are not limited to, replica rejuvenation~\cite{bessani2008crutial, garcia2019lazarus}, diversity~\cite{bessani2008crutial, garcia2019lazarus, chun2008diverse, distler2011spare}, and N-version programming~\cite{distler2011spare}. 
To enhance adaptability further, an ITS can integrate information from public sources such as social networks, news outlets, and online databases that track vulnerabilities, exploits, and cyber threats. 
This integration enables the ITS to proactively adjust its defenses by modifying configurations and operating systems (OSs), thereby increasing resilience against identified threats.

Risk assessment has long been a cornerstone of cybersecurity research, evolving through various approaches to address emerging threats.
Early studies focused on leveraging information from NIST’s database, particularly CVSS scores, to systematically evaluate system vulnerabilities~\cite{houmb2008estimating}.
Building on this foundation, subsequent research incorporated data from public social networks to improve situational awareness and provide richer contextual insights~\cite{alves2020follow}.
As technology advanced, automated methods emerged to streamline risk assessments, particularly for operating systems, by analyzing associated CVEs~\cite{heo2017designing}.
Complementing these efforts, other investigations explored the role of system diversity in minimizing risk, highlighting how the quantity and nature of CVEs influence exploitability~\cite{garcia2011diversity}.
Further refinements in risk assessment have introduced alternative CVSS scoring mechanisms for CVEs, reusing existing data to enhance evaluation precision~\cite{garcia2019lazarus}. 
These contributions collectively highlight the importance of integrating diverse data sources and methodologies to strengthen proactive and adaptive risk management strategies.
These solutions leverage online exploit and vulnerability databases to gather relevant information, assess the associated risks to the current system using base scores from the retrieved data, and recommend lower-risk alternatives when feasible. 
However, a notable trade-off arises from the reliance on these base scores, which require manual evaluation of each vulnerability before assignment.
This process introduces a significant delay between discovering new exploits or vulnerabilities, their subsequent risk assessment, and the overall evaluation of the ITS~\cite{ruohonen2019look}.

The process for a new Common Vulnerabilities and Exposures (CVE) creation begins when a vulnerability is submitted for evaluation to the NVD.
Initially, it is added to the CVE database, accompanied solely by the description provided by the applicant~\cite{cveprocess}.
Once listed in the CVE database, the vulnerability undergoes the CVE Analysis process, providing additional details, including contextual information and preliminary assessments.
The evaluation process involves several key steps to ensure the accuracy and utility of the CVE record.
First, an analyst verifies the references provided with the CVE submission and assigns appropriate reference tags to facilitate data discovery. 
Then, it manually searches for additional public information relevant to the CVE.
During this process, a Common Weakness Enumeration (CWE) identifier is assigned to specify the type of vulnerability associated with the CVE.
Subsequently, the Common Vulnerability Scoring System (CVSS) version 3.1 metrics for exploitability and impact are calculated based on the gathered public information and the guidelines outlined in the submission.
The analyst then prepares a Common Platform Enumerator (CPE) Applicability Statement, which identifies all software and/or hardware potentially affected by the vulnerability. 
Finally, before the CVE is published, a senior analyst conducts a comprehensive review and quality assurance check to ensure the record's accuracy and completeness.

Because of these steps, the evaluation process is inherently lengthy.
This creates a time gap between the reception and evaluation of a vulnerability, which usually results in a delay that is often proportional to the severity of the vulnerability. 
For particularly critical vulnerabilities, the evaluation process can extend for as long as a year~\cite{ruohonen2019look}.
Moreover, as of the time of writing, the backlog of vulnerabilities in the NVD has been significantly increasing due to a lack of sufficient resources for processing new CVEs.
For instance, in March 2024, only 199 of the 3,370 CVEs submitted to the NVD were successfully analyzed~\cite{nvdbacklog}.
If this issue remains unresolved, the response time for addressing newly reported vulnerabilities is likely to grow further, jeopardizing the reliability of the NVD CVE database as a trusted and reliable source of information.

To address these challenges, this work builds upon and extends the research presented in ``HAL 9000: A Risk Manager for ITSs'', accepted at \textit{The Sixth IEEE International Conference on Trust, Privacy, and Security in Intelligent Systems and Applications}.
The original paper proposed secure and resilient configurations for ITSs by leveraging publicly available Open-Source Intelligence (OSINT) data and automating the assessment of CVSS scores for unrated CVEs.
Recognizing the growing volume of information in publicly available OSINT databases, this extension enhances the original framework by integrating additional sources of information and refining the calculation step for risk assessment. 
These advancements aim to improve the system's accuracy and depth in evaluating risks, further enhancing its ability to predict CVSS scores for unrated CVEs and dynamically reassess ITS configurations.

A review of the state-of-the-art reveals that current Risk Managers~\cite{heo2017designing, garcia2019lazarus} rely heavily on the NVD's CVE evaluation process, as their risk calculations are based on the CVSS base scores provided by the NVD and the existence of an exploit. 
HAL's approach overcomes this limitation, offering a more proactive and efficient response to emerging threats.
HAL's CVSS score prediction tool aims to provide a temporary score, enabling faster response and minimizing the vulnerability window against emerging threats, thereby enhancing the adaptability of ITSs.
This tool is not intended to permanently replace the NVD evaluation process, as specific nuances and subtleties in vulnerabilities can only be accurately identified through human analysis.
Once the NVD evaluation is complete, HAL updates the scores of the previously predicted CVEs and re-executes the configuration risk assessment function, ensuring that its risk evaluations remain accurate and aligned with the most reliable information available.

HAL's primary objective is to recommend secure configurations for ITSs using knowledge obtained from publicly available OSINT databases. 
However, given the high-risk environments in which ITSs are deployed, it is critical to ensure HAL's resilience against potential malicious attacks. 
A secure deployment environment for HAL is strongly recommended to achieve this.
For instance, both Lazarus~\cite{garcia2019lazarus} and Skynet~\cite{freitas2023skynet} adopt a two-plane architecture to enhance security. 
This architecture separates the system into a secure, isolated controller plane that manages critical operations and an execution plane that interfaces with the environment and is exposed to potential threats. 
Adopting a similar approach for HAL can significantly bolster its robustness and ensure reliable operation in adversarial conditions.
This concept is already established within the Software Defined Networks (SDN) research, where the SDN controller is separated from the SDN switches used in the data plane~\cite{bannour2017distributed}.

The primary goal of deploying a Risk Manager in a secure environment is to ensure its correct operation remains uncompromised by malicious adversaries. 
Such attacks could target the system in multiple ways, including tampering with the local OSINT database—such as HAL’s—via input injection~\cite{malik2016database} or compromising the ML model through data poisoning~\cite{yerlikaya2022data}.
To mitigate these risks, it is assumed that ITSs employing Risk Managers can provide a secure execution environment, safeguarding both the data sources and the integrity of the ML models.
This foundational security measure is critical to maintaining the reliability and effectiveness of the Risk Manager in dynamic and adversarial settings.
In summary, this paper extends the capabilities of the published Risk Manager, HAL 9000, through the following key advancements:
\begin{itemize} 
    \item Advancing the state-of-the-art in Risk Management by enhancing existing methodologies and integrating new approaches. 
    \item Providing a comprehensive description of the scraper developed to collect exploit and vulnerability data from diverse OSINT platforms and other publicly available sources.
    \item Outlining the process of assembling the experimental dataset, ensuring transparency and reproducibility.
    \item Reassessing and refining the method for calculating the risk associated with an OS, improving the accuracy and reliability of the risk evaluation. 
\end{itemize}

These contributions collectively strengthen the effectiveness and applicability of HAL 9000 in managing risks within ITSs.

The rest of the paper is organized as follows:
Section~\ref{sec:related_work} reviews the current state of the art in Risk Managers, CVSS score predictors, text document clustering algorithms~\cite{singhal2017survey}, and OSINT scrapers.
Section~\ref{sec:design} provides a detailed description of HAL's architecture and workflow.
Section~\ref{sec:performance} presents the experimental setup and a comparative discussion of the results.
Finally, Section~\ref{sec:conclusion} discusses the main conclusions of this research and offers perspectives for future work.

\section{Related Work}\label{sec:related_work}
The present section addresses the current state of the art in Risk Managers, emphasizing their roles in evaluating system configurations and providing recommendations for secure and resilient setups.
Additionally, it reviews proposed solutions for CVE score prediction methodologies and the application of machine learning techniques, particularly for clustering text documents~\cite{hotho2005brief}.

\subsection{Risk Managers} 

\noindent\textbf{Diversity Policy for Intrusion Tolerant Systems}~\cite{heo2017designing}, proposed a diversity policy to be employed by recovery-based ITSs. 
It retrieves information on known software vulnerabilities and advises on combinations that minimize the risk of common vulnerabilities.
This approach generates all possible configurations and decides based on the minimum sum of the CVSS scores.

This architecture is designed for web servers that deliver services accessible via the Internet.
It consists of a cleansing group, a central controller, and a processing group. 
The cleansing group hosts Virtual Machines (VMs) as they undergo recovery, after which the restored VMs migrate to the processing group, providing the designated web service and facing external threats. 
The central controller coordinates the allocation and lifecycle of the VMs across both groups and executes the configuration selection algorithm to determine the optimal system setup.
The system applies a proactive recovery mechanism to protect the system from undetected attacks. 
A central controller periodically rotates the state of each VM, transitioning between active, cleansing, ready, and active states. 
During the transition from the cleansing to the ready state, the central controller executes a selection algorithm, applying the diversity policy to decide on a suitable secure configuration.

Yet, the drawback of employing the diversity policy method lies in its dependency on the NVD database score before implementing any system alterations. 
This reliance can result in delays in adapting to the new vulnerabilities~\cite{ruohonen2019look} because they were added to the CVE database and have not been evaluated by NVD, i.e., do not have a score attributed.\\

\noindent\textbf{Lazarus}~\cite{garcia2019lazarus} introduced a Byzantine Fault Tolerant (BFT) architecture that adjusts system configurations to reduce vulnerability risks based on recommendations from its Risk Manager module. 
The research proposes a novel CVE scoring method.
It uses ML algorithms to detect shared CVEs\footnote{By Lazarus definition, a CVE is shared between two or more OSs by the CVE vulnerability configuration field and/or by the description similarity with a CVE from another OS.} between replicas using the CVE description.
The new calculation method reevaluates the CVSS base score using the information in the CVE details.
This includes factors such as the age of the CVE, the existence of a patch, and the availability of an exploit, as presented in the following equations.

\begin{equation}\label{eq:1}
 score(v) = CVSS(v) \times oldness(v) \times exploited(v) \times patched(v)
\end{equation}


\begin{equation}
\begin{split}
oldness(v) = max \Biggl( 1 - 0.25 \times \frac{(now - v.published\_date)}{oldness\_threshold}, 0.75 \Biggr)
\end{split}
\end{equation}

\begin{equation}
patched(v) = 0.5^{v.patched}
\end{equation}

\begin{equation}\label{eq:final}
patched(v) = 1.25^{v.exploited}    
\end{equation}

Another contribution of Lazarus was its ability to identify different CVEs in the NVD database that describe similar vulnerabilities affecting distinct applications or operating systems (OSs).
This was achieved through description analysis.
For example, CVE-2014-0157 affects OpenSUSE 13, CVE-2015-3988 affects Solaris 11.2, and CVE-2016-4428 affects Debian 8.0. 
Despite targeting different OSs, their descriptions share similarities, indicating a potential common exploit across these systems.
To analyze CVEs that follow the same pattern, Lazarus employed K-means, a clustering technique from ML, to verify their description.
K-means is an algorithm for unsupervised learning that clusters the CVEs by their description similarity.
After clustering, in the evaluation phase, Lazarus's Risk Manager employs the new scoring system, penalizing pairs of replicas with common CVEs or the same clustered CVEs.
Clustering suggests a likelihood of the exploit impacting both replicas.
Based on this evaluation, Lazarus recommends a more resilient system configuration against intrusions.

The Risk Manager used in Lazarus introduced a trade-off between security and resilience. 
It highlights that a system solely focused on minimizing risk may not be ideal for BFT protocols. 
When a pair of replicas can have one or more shared unpatched CVEs, it is possible to compromise the system by executing parallel attacks.
Even though the recommended configuration might not be the most secure, there is a guarantee that an unpatched vulnerability will not affect more than one replica simultaneously, maintaining the BFT invariant of $3f + 1$.
Besides the drawback of being dependent on the NVD score system, Lazarus's Risk Manager also requires the human intervention necessary to execute the K-means algorithm. 
The optimal number of generated clusters has to be manually visualized and inserted into the algorithm, which increases the execution time and, subsequently, the vulnerability time window.
In addition, K-means is an algorithm susceptible to outliers~\cite{singh2013analysis}, i.e., the mean value of a cluster can be influenced by the outliers, which will affect the resulting clusters.
This can lead to wrong assessments.

\subsection{CVSS score prediction}\label{subsec:score_prediction}

With the increasing applications of Artificial Intelligence (AI) and the latest advancements in Deep Learning, computers can now automate or replicate many processes. 
AI and Deep Learning algorithms can replicate tasks that are manually performed repeatedly and follow consistent patterns. 
This automation can speed up these processes, allowing operators to focus on refining and verifying the results and providing feedback to improve the algorithms.

Given the lengthy evaluation process for new CVEs and the challenges evaluators face, researchers have explored new procedures to apply AI and Deep Learning algorithms that can automatically predict the CVSS scores for CVEs, enabling a quicker assessment.
As such, the related work on CVSS score predictors is addressed.\\

\noindent\textbf{Khazaei et al.~\cite{khazaei2016automatic}} introduced a method for predicting CVSS scores using natural language processing (NLP).
This method learns from previously available CVE vulnerabilities by analyzing their descriptions and associated CVSS scores.
The method applied text mining tools and techniques to extract feature vectors during data preprocessing. 
The authors evaluated the application of three different algorithms for predicting CVSS scores: Support Vector Machines (SVM), Random Forest, and fuzzy systems.
Among these, the fuzzy systems provided the best accuracy, correctly scoring 88\% of the evaluated CVEs.
In conclusion, the study found that using automatic predictors can reduce human error and increase the speed of CVSS score calculation.\\

\noindent\textbf{Sahin et al.}\cite{sahin2019conceptual} extended the prior work of Han et al.\cite{han2017learning}, which predicted the severity levels of vulnerabilities based on their descriptions. 
Their study used similar feature extraction methods with word embeddings and prediction models using Convolutional Neural Networks (CNNs). 
Additionally, it incorporated Long Short-Term Memory (LSTM) networks and Extreme Gradient Boosting (XGBoost).

The objective of this research was to predict severity scores in addition to severity levels. 
The original work categorized vulnerability severity into ranges: low (0.1 - 3.9), medium (4.0 - 6.9), high (7.0 - 8.9), and critical (9.0 - 10.0).
For feature extraction, the authors removed words that only occurred once in the sentence corpus and trained the word2vec continuous skip-gram model, introduced by Mikolov et al.~\cite{mikolov2013efficient}. 
After generating feature vectors for each word, they converted the description sentences into vector representations by concatenating the word vectors.
The authors concluded that vulnerability descriptions contain valuable information that can be used with deep learning algorithms such as LSTM, CNN, and gradient boosting to predict severity scores with an average error of 16\%.\\

\noindent\textbf{Elbaz et al.~\cite{elbaz2020fighting}} proposed using CVSS vector prediction to address N-Day vulnerabilities.
CVSS vector prediction aims to forecast the base metrics that constitute the CVSS standard and assess the severity of vulnerabilities. 
This approach uses linear regression on vulnerability descriptions to provide basic metrics for newly discovered vulnerabilities.

This method differs from previous work as it predicts the individual metrics rather than the overall score, allowing CVSS to offer more detailed information about the vulnerability and enhance the  ``explicability'' of the results.
The authors used a bag-of-words approach to the vulnerability descriptions and a filtering scheme to remove irrelevant words. 
Subsequently, a regression model was trained for each metric to be applied during the assessment.
The authors concluded that while the method by Khazaei et al.~\cite{khazaei2016automatic} achieves higher accuracy for the base score, their approach offers better explainability.
They recommend implementing two pipelines: one for accuracy prediction and the other for ``explicability''.\\

\noindent\textbf{Costa et al.\cite{costa2022predicting}} proposed combining text preprocessing using NLP techniques with vocabulary expansion and the application of the Deep Learning method DistilBERT\cite{sanh2019distilbert}.
The objective of the research was to predict CVSS metrics using vulnerability descriptions, similar to the work of Elbaz et al.~\cite{elbaz2020fighting}.
For preprocessing, lemmatization and stemming were applied to the data.
Tokenization was performed using the Transformers library. 
The experiments tested the accuracy of combining the data with vocabularies of 5,000, 10,000, and 25,000 words added to the tokenizer's vocabulary.
The authors then tested several deep learning methods, specifically DeBERTa, BERT, ALBERT, and DistilBERT, to evaluate the accuracy and quality of the assessed data. 
The results showed that DistilBERT provided the most accurate predictions, while ALBERT was the least accurate.
The study concluded that DistilBERT is a state-of-the-art model for CVSS prediction, with enhanced performance when combined with lemmatization and a 5,000-word vocabulary.
However, no analysis was made on the accuracy of the base score since the main idea is to provide insight for the experts.\\

\noindent\textbf{Kai et al.\cite{kai2023vuldistilbert}} developed VultDistilBERT, a method to assess vulnerability severity, similar to the approach by Sahin et al.\cite{sahin2019conceptual}, using a distillation model. 
The technique addresses data imbalance by augmenting data and using optimal subsets.
The data augmentation process generates more raw data without increasing quantity. This is achieved by synonym replacement and random deletion to expand the sample set. The optimal subset selection involves choosing a subset of CVSS metrics during training to incorporate into the vulnerability descriptions, thereby enhancing textual information.
The method then uses the DistilBERT model as a text characterization tool, processing the preprocessed data. The resulting feature vectors from DistilBERT are then classified using a linear layer.
The authors concluded that their proposed method achieved state-of-the-art performance in vulnerability severity assessment, with a 97\% assessment accuracy.

\subsection{Text Document Clustering}\label{subsec:clustering}

Text clustering is a technique used in text mining and information retrieval.
Initially investigated to enhance the precision or recall in information retrieval systems~\cite{rijsbergen1979information, kowalski2007information}, as an efficient way of finding the nearest neighbors of a document~\cite{buckley1985optimization}, in browsing a collection of documents~\cite{cutting2017scatter}, or in organizing the results of a search engine response~\cite{zamir1997fast}.

Text Document Clustering is an unsupervised technique that groups documents into categories using unsupervised learning algorithms. 
According to Lazarus, online databases may contain entries with different identifiers that affect distinct systems but have identical descriptions, indicating the same vulnerability.
Applying text document clustering can mitigate this inaccuracy, enhancing the quality of data collected from online vulnerability databases. 
Therefore, the present subsection addresses the related work on Text Document Clustering.\\

\noindent\textbf{Steinbach et al.}~\cite{steinbach2000comparison} compared the results of applying agglomerative hierarchical clustering with K-means for document clustering. 
The study evaluated the implementations of the Unweighted Pair Group Method with Arithmetic Mean (UPGMA), K-means, and bisecting K-means.
To measure the quality of these algorithms, the authors used an internal quality measure, overall similarity, and an external quality measure, entropy. 
Overall similarity assesses cluster cohesiveness, while entropy measures the quality of the created clusters.
The authors concluded that ``given the linear run-time performance of bisecting K-means and the consistently good quality of the clusterings that it produces, bisecting K-means is an excellent algorithm for clustering a large number of documents''~\cite{steinbach2000comparison}.\\

\noindent\textbf{Zhao et al.}\cite{zhao2002comparison} compared the application of Agglomerative algorithms with Partitional algorithms. 
They evaluated group average functions against \textit{k}-way clustering solutions. 
To assess the results from each algorithm, they used the FScore, introduced by Larsen et al.\cite{larsen1999fast}, and entropy measures.

A perfect clustering solution would produce clusters containing documents from only a single class, resulting in zero entropy. Generally, lower entropy values indicate better clustering solutions. 
The FScore is necessary to account for clustering solutions that may seem poor due to the presence of outlier documents. 
Higher FScore values indicate better clustering solutions.
Their results concluded that \textit{k}-way clustering solutions outperformed those from agglomerative clustering implementations.\\

\noindent\textbf{Singh et al.}~\cite{singh2011document} compared the performance of K-means, Heuristic K-means, and Fuzzy C-means for document clustering. 
The study explored different feature selection conditions (with and without stop words, with or without stemming) and various representations (term frequency, term frequency-inverse document frequency, and Boolean).
The trade-offs between the algorithms are as follows: K-means clustering quality is sensitive to initial seeds, Heuristic K-means adds computational cost due to the use of heuristics, and Fuzzy C-means does not produce hard clusters but provides a degree of membership for all created clusters.
The authors used internal and external criteria for evaluation: residual sum of squares and purity, respectively. 
The results showed that Heuristic K-means performs better than K-means, but Fuzzy C-means is a more robust flat clustering algorithm.\\

\noindent\textbf{Mendonça et al.}\cite{mendoncca2019clustering} studied the effectiveness of classical literature clustering algorithms when applied to free text documents. 
They selected five clustering algorithms for their experiments, each capable of word-embedding document representation: K-means\cite{hartigan1975clustering}, Expectation–Maximization (EM) Clustering using Gaussian Mixture Models (GMM)\cite{liu2002document}, Spectral Clustering\cite{shi2000normalized}, Mean Shift~\cite{comaniciu2002mean}, and Density-Based Spatial Clustering of Applications with Noise (DBSCAN)~\cite{ester1996density}. 
The study aimed to observe the behavior of these algorithms in the task of document clustering.
For evaluation, the authors used Normalized Mutual Information and Homogeneity scores to measure the overall quality of the clustering solutions.
The results indicated that the chosen parameters significantly influence the performance of the algorithms. 
Among the selected algorithms, K-means performed the best.\\

\noindent\textbf{Asyaky et al.}\cite{asyaky2021improving} conducted an experiment to enhance the performance of density-based algorithms, specifically DBSCAN\cite{ester1996density} and HDBSCAN~\cite{mcinnes2017accelerated}.
The authors preprocessed the documents using lemmatization, stemming, and document embeddings to reduce dimensionality and improve the accuracy of these clustering algorithms.
For evaluation, they used the Adjusted Rand Index, which compares the pairs of objects in the resulting clustering to a ground truth clustering, and the Adjusted Mutual Information, which is based on information-theoretical mutual information~\cite{wagner2007comparing}.
The authors concluded that the obtained results surpassed most existing methods on the state of the art on the same subject.

\subsection{OSINT Scraper}
Given the extensive volume of OSINT data available for cybersecurity, vulnerabilities, and exploits, systems that rely on these sources require mechanisms for continuous data retrieval.
Manual input is highly resource-intensive and may not capture all relevant information.
To address this challenge, developers frequently employ web scrapping\footnote{Whereas web crawlers traverse the Internet in search of new sites relevant to a topic, scrapers typically target already identified websites to extract specific information.}, which automates the process of systematically and rapidly collecting critical data—such as indicators of compromise, exploit discussions, and newly discovered vulnerabilities. 
In the present subsection, we review the state-of-the-art OSINT scrapers.\\

\noindent\textbf{Fernandes et al.}~\cite{fernandes2023scrapeioc} proposed ScrapeIOC, a web-scraping tool specifically designed to retrieve publicly available Indicators of Compromise (IOCs).
The main objective was to collect as many IOCs as possible from various online sources, targeting specific types and families of malware. 
Leveraging ScrapeIOC, the authors constructed an offline database that organizes IOCs—primarily hashed samples (MD5, SHA1, or SHA256)—by their source.
They then integrated this database with publicly accessible repositories to create a search engine for malware hashes. 
The tool scrapes multiple platforms, including Malwares, Malshare, VxCube, ThreatCrowd, and Maltiverse.
Although ScrapeIOC effectively gathers critical IOCs, it is specialized for retrieving malware-related hashes rather than offering broader intelligence, such as vulnerabilities, exploits, or proof-of-concept details.
Consequently, while it addresses an important facet of cyber defense, ScrapeIOC does not provide the more generalized capabilities found in frameworks like HAL, which support a wider range of cybersecurity intelligence.\\

\noindent\textbf{Alves et al.}~\cite{alves2021processing} introduced SYNAPSE, a tool designed for extracting security-related information from Twitter\footnote{Twitter and the Twitter logo are now officially rebranded as X since 2023.} accounts belonging to individual users, security organizations, and researchers. 
Its goal is to provide a continuous threat monitor that synthesizes and summarizes relevant intelligence for security analysts. 
The SYNAPSE pipeline encompasses filtering, feature extraction, binary analysis, clustering, and the generation of IOCs. 
Data collection begins with a scraping mechanism based on Twitter’s publicly available API, wherein selected accounts are chosen based on their likelihood to share security-specific content.
However, this approach confines the collected information to Twitter, excluding insights from other potentially valuable data sources.\\

\noindent\textbf{Kühn et al.}~\cite{kuehn2023common} introduced a classification scheme for the suitability and crawlability of reference texts in the NVD.
Their research aimed to predict CVSS vectors using Deep Learning techniques.
To achieve this, they developed a web scraper to retrieve vulnerability-related texts from external websites referenced by the NVD.
The authors categorized these referenced websites into several groups based on their characteristics, including version control and bug tracker services, mailing lists, patch notes, security advisories, third-party articles, and blogs/social media.
Since most of these websites did not provide APIs for data retrieval, the authors implemented a custom scraper tool to collect the necessary information.
However, the tool was specifically designed for scraping websites referenced by the NVD and did not consider other publicly available sources that may also contain valuable information.

\section{HAL 9000 Architecture}\label{sec:design}

This section presents an overview of HAL 9000's extended architecture and workflow and a comprehensive description of the data scraper developed to extract information from online OSINT databases and public sources.
Building on the findings of Pastor et al.~\cite{pastor2020not}, which highlight the scarcity of tools for OSINT information retrieval, this section also details the implementation process for the scraper employed.

\begin{figure}[htbp]
  \centering
  \includegraphics[width=0.2\linewidth]{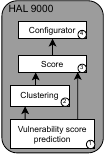}
  \caption{HAL 9000 Architecture and workflow.}
  \label{fig:arch}
\end{figure}

\noindent\textbf{HAL's architecture}: depicted in Figure~\ref{fig:arch}, is constituted by four components: the Vulnerability score predictor, the Clustering algorithm, the Score reassessment, and the Configurator.

The Vulnerability score predictor assesses new CVEs by providing a score using its description.
It uses an ML model trained with established vulnerability data. 
Specifically, it uses the information in the CVE's description and CVSS score metrics to train the model.
Afterward, it assesses the new CVEs through their description and/or CVSS metrics depending on the applied algorithm, predicting the CVSS base score.
The predicted CVSS score is then reassessed using the NVD equations~\cite{calculator}.

HAL, in its implementation, based on initial studies, used the algorithm proposed by Khazaei et al.~\cite{khazaei2016automatic} to accomplish this objective.
However, a modular approach was adopted during the architecture design to facilitate interchange between different algorithms.
This was achieved by restricting interaction between elements to the bare minimum, i.e., the dataset is transmitted to the element for training, thereafter accepting only the data to be evaluated.
Both datasets are passed in a CSV file containing the CVE ID, description, and CVSS metrics.

\begin{figure}[htbp]
    \centering
    \includegraphics[width=0.7\textwidth]{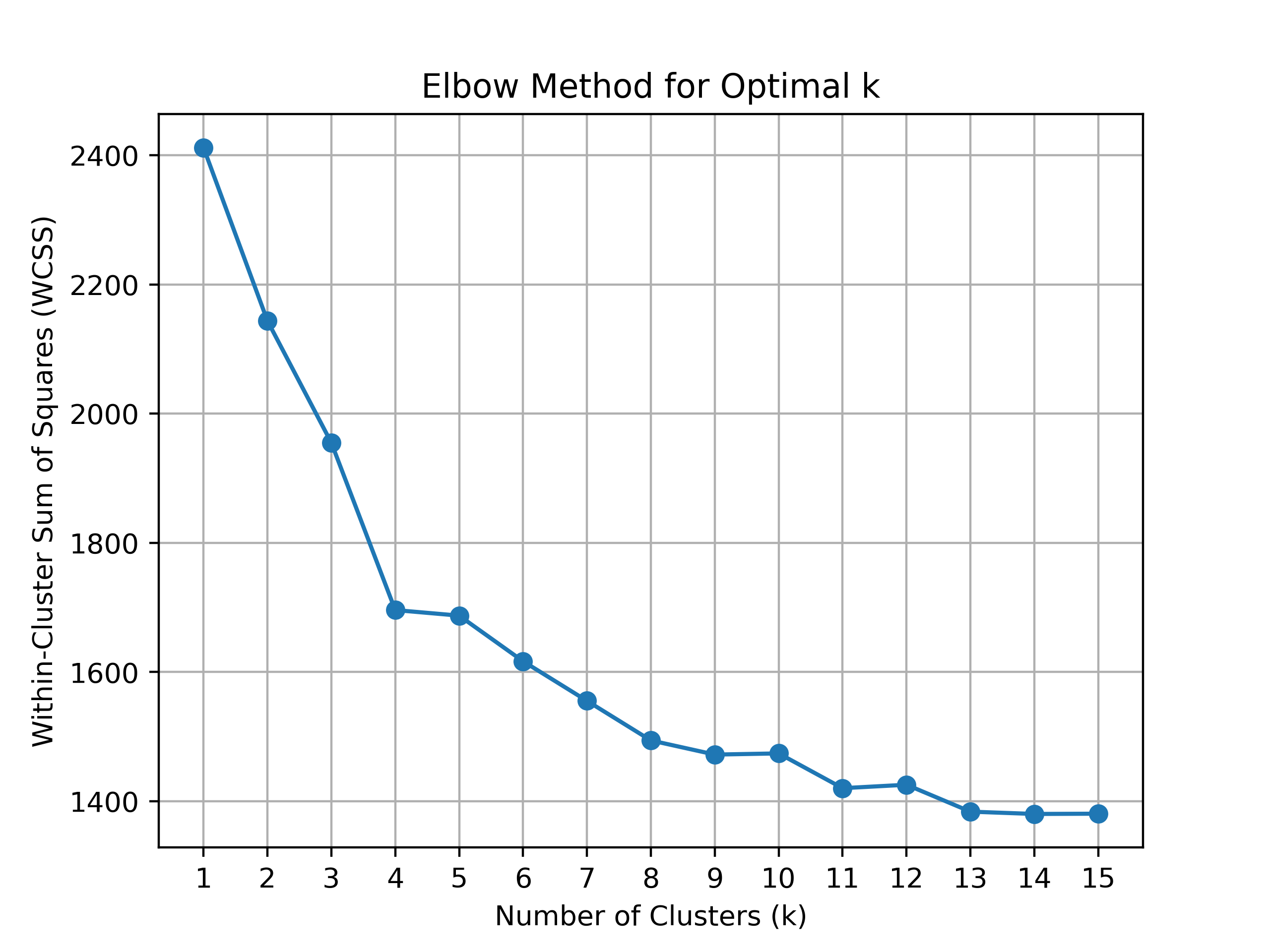}
    \caption{Example of a graph where visual extracting K can be challenging. According to the Pelt~\cite{killick2012optimal} method, the best K is 12.}
    \label{fig:elbow_method}
\end{figure}

The Clustering algorithm applied in the architecture follows the concept established by Lazarus.
Databases can have different CVE entries that affect different software/OSs but are similar in vulnerability.
This component aims to identify these occurrences through clustering using the CVE description.
Clustering is done through the use of an ML algorithm. 
This creates a set of clusters to classify each description based on their similarity.
For this component, the ML algorithm should be fully automated to reduce the total execution time window and eliminate the possibility of human bias or error.
Figure~\ref{fig:elbow_method} shows an example of the graph generated during the Lazarus execution to apply the Elbow method, i.e., visually decide on the number of clusters to be used.
Furthermore, the algorithm must be robust to outlier data and prevent inaccurate clustering when inferring new data. 
Examples of ML algorithms that demonstrate insensitivity to outliers include, but are not limited to, HDBSCAN~\cite{mcinnes2017accelerated} and OPTICS~\cite{scikit-learn}.
Nonetheless, HAL's architecture was designed to accommodate other ML algorithms, i.e., supervised, semi-supervised, and deep learning. 
The component follows the same modular design as the Vulnerability score predictor to allow this.

The Score reassessment component extends Lazarus's contribution.
It recalculates the CVSS score to reflect the CVE vulnerability more accurately.
As mentioned in section~\ref{sec:related_work}, Lazarus recalculates the score using the additional information in the CVE to improve its quality by taking into account its age, patch availability, and occurrence of an exploit~\footnote{A vulnerability is considered exploited when there are reports of its occurrence or when code is provided to exploit the vulnerability.}.
HAL's architecture augments the Lazarus scoring system by considering the attention that the vulnerability receives in the ``wild''.
Lazarus equations make minor adjustments to the CVSS base score, halving the score when a patch is available.
However, in practical terms, a patch does not guarantee its installation.
In 2022, incident responders were brought in to remediate attacks that began with exploited vulnerabilities, such as the ProxyShell and Log4Shell vulnerabilities. 
Both had existing patches at the time of compromise~\cite{unpatched}.

In its original formulation, HAL extended the calculation methodology of Lazarus by readjusting the score to account for the potential non-installation of a patch and by incorporating the weight of the Exploit Prediction Scoring System (EPSS)~\cite{jacobs2021exploit}, as shown in equation\ref{eq:hal}.
The EPSS quantifies the probability of a CVE being exploited in the wild in the next 30 days.
Subsequently, HAL computes the \textit{hal\_score(v)}, as presented in equation~\ref{eq:final}.
This computation involves determining the \textit{score(v)} both with and without the application of a patch (if available) and then applying the weighted probabilities derived from the EPSS for scenarios where the CVE is either targeted or not targeted.
When a patch is unavailable, the corresponding terms in the \textit{hal\_score(v)} equation that accounts for patch-related adjustments are assumed to be zero.
The current extension enhances the score reassessment by considering the presence of CVEs in pulses\footnote{\textbf{Pulses}, as defined by AlienVault, are curated collections of threat intelligence information shared by users or organizations. 
These collections contain actionable data about specific threats, including associated IOCs, descriptions, tags, related vulnerabilities, and techniques employed by attackers.} from AlienVault Open Threat eXchange (OTX). 
Pulses provide evidence that a CVE is actively exploited and is used across diverse attack vectors, indicating its broader exploitation potential. 
The calculation is refined using curated threat intelligence information from AlienVault to include the CVE's occurrence in distinct pulses, presented in equation~\ref{eq:hal_new}.

For example, CVE-2017-11882 is an exploit with a CVSS score of 7.8, classified as "High" under NVD standards, and has a patch available. 
According to the Lazarus calculation, this vulnerability is reassessed to a score of 3.65, which falls into the "Low" category as per the Qualitative Severity Rating Scale~\cite{qualitative}. 
However, this reassessment does not accurately reflect the exploit's risk profile.
Discovered in 2017, CVE-2017-11882 remains an active threat despite the availability of a patch, with an exploit probability of 97.99\% as derived from EPSS data. Using the methodology proposed in HAL's previous work, the recalculated score was adjusted to 7.2, maintaining its "High" classification. This adjustment provided a more realistic evaluation, accounting for scenarios where patch application might be delayed.
However, this prior approach still failed to capture this vulnerability's severity fully, given its presence in 50 distinct Pulses.
With the proposed extension in the current work, the recalculated score rises to 8.9, placing it on the borderline of the "Critical" category. 
This adjustment more accurately reflects the actual risk posed by CVE-2017-11882, highlighting the urgency of its mitigation.

\begin{equation}\label{eq:new}
\textrm{score}(v) = \textrm{CVSS}(v) \times \textrm{oldness}(v) \times \textrm{exploited}(v) \times \textrm{patched}(v)
\end{equation}


\begin{equation}\label{eq:hal}
\textrm{hal\_score}(v) = \textrm{score}(v) \times ( 1 - \textrm{EPSS}(v)) + \textrm{score}(v)_{wp} \times EPSS(v) 
\end{equation}

\begin{equation}\label{eq:hal_new}
\textrm{hal\_score}(v) = \min\Bigg(10, \textrm{score}(v) \times \big( 1 - \textrm{EPSS}(v) \big) + \textrm{score}(v)_{wp} \times \textrm{EPSS}(v) + \log(\#\textrm{related\_pulses})\Bigg)
\end{equation}

The final component, the Configurator, uses the gathered information to propose a secure and resilient configuration for the ITS to deploy.
The component outputs two values, the \textit{security\_risk(config)} and the \textit{resilient\_risk(config)}.
Its objective is to minimize both outputs, prioritizing the \textit{resilient\_risk(config)}, i.e., shared CVEs calculation to avoid situations of ``break one, break all'', where parallel attacks target ITS nodes.
The \textit{security\_risk(config)} is calculated by summing the newly assessed CVSS scores and predicted scores for new CVEs in a given configuration from all participating nodes.
In the resilience calculation, the algorithm pairs every node in the configuration two by two and sums the CVSS score between shared CVEs and between CVEs clustered by their similarity.

\begin{equation}\label{eq:security}
security\_risk(config) = \sum_{n_i \in config} \sum_{v \in V(n_i)} hal\_score(v)
\end{equation}

\begin{equation}\label{eq:resilience}
resilience\_risk(config) = \sum_{n_i, n_j \in config} \sum_{v \in V(n_i, n_j)} hal\_score(v)
\end{equation}

\subsection{HAL 9000 workflow}
HAL's workflow, depicted in Figure~\ref{fig:arch}, illustrates the data path from OSINT data retrieval to the recommended configuration.
This subsection provides detailed information on HAL's workflow, describing the different paths the data takes through its various components during execution.

HAL's workflow is divided into a four-step process:
\begin{enumerate}
    \item Predicting CVE scores for unassessed CVEs.
    \item Clustering similar CVEs based on their descriptions.
    \item Performing a risk assessment of the system configuration.
    \item Providing the most resilient and secure configuration.
\end{enumerate}

\noindent\textbf{1st step:} HAL receives a dataset of vulnerabilities and identifies CVEs that do not have assigned scores, i.e., CVEs that only have descriptions, with status ``Received''. 
HAL uses an AI score predictor model to estimate the scores based on the evaluation patterns observed in the assessed vulnerabilities from the training dataset for these unassessed vulnerabilities.\\

\noindent\textbf{2nd step:} The CVEs are clustered based on their descriptions to identify similar vulnerabilities. 
This process avoids situations where different software is affected by distinct vulnerabilities, but it is the same vulnerability, considering that both descriptions are the same.\\

\noindent\textbf{3rd step:} HAL then reassesses the scores of the vulnerabilities using previously established equations (refer to equations~\ref{eq:1} to~\ref{eq:hal}). 
This process is also applied to CVEs assessed by the Vulnerability Score predictor. 
New CVEs have the necessary information to apply the new equations to the predicted score, specifically, the added date, if they were exploited, and the existence of a patch\footnote{Newly added vulnerabilities typically do not have patches for application}.\\

\noindent\textbf{4th step:} After reassessing the scores of CVEs, HAL generates all conceivable variations of the given software/Operating System (OS) configurations. 
HAL calculates its security and resilience levels for each configuration using equations~\ref{eq:security} and~\ref{eq:resilience}, respectively. 
The resulting set of configurations is then arranged in order of resilience and security levels. The top configuration, representing the most resilient and secure option, is recommended to the ITS.\\

\subsection{Open-source Intelligence scraper}

A scraper was developed for OSINT databases and other intelligence sources, including news, social networks, and blogs, to complement HAL's implementation.
Given the continuous growth in vulnerabilities and exploits, an automated scraper benefits the system by continuously retrieving new vulnerabilities from online sources.
However, directly gathering information from online OSINT databases and other sources could cause delays in HAL's execution. 
Therefore, the optimal solution was establishing a local database where the scraper could write and update the existing information.

Pastor-Galindo et al.\cite{pastor2020not} demonstrated that research for OSINT collection and analysis has yet to expand, as evidenced by the limited number of available open-source solutions.
For example, current solutions like searchsploit\cite{searchsploit} only provide information from a single database, ExploitDB. 
This subsection provides a detailed description of the OSINT scraper and its development process to guide future implementations.\\

The scraper was designed to query four different sources of OSINT databases/information every hour: the NVD CVE database~\cite{booth2013national}, ExploitDB~\cite{exploitdb}, AlienVault OTX~\cite{alienvault}, and the Open Source Vulnerabilities (OSV)~\cite{osv_dev} database.
To access the AlienVault OTX and the NVD CVE database, an Application Programming Interface (API) key was required due to rate limits imposed by the database\footnote{50 requests per 30 seconds (with key), five requests per 30 seconds (without key)~\cite{developersNVD}.}. 
This measure prevents denial of service attacks from malicious bots. 
If the limits are exceeded, the database becomes unresponsive for that IP address. 
Therefore, the scraper uses API keys provided by the OSINT databases and adheres to these limits to ensure queries remain within specified bounds.
Accessing NVD database information requires forming REST-compliant requests following NVD's specifications~\cite{nvdapi}. 
The NVD's response comes in JSON format, which has to be parsed to retrieve the information.
The number of CVEs in a response is limited to 2,000 entries to manage data flow. 
Queries exceeding this limit require additional requests for subsequent pages, as specified by NVD~\cite{nvdapi}. 
Additionally, the database may delay responses for consecutive queries from the same IP address to prevent resource drainage.\\

For ExploitDB's information, it was impossible to create a scraper due to protections against web scraping and crawling.
Attempts to remotely access the database are blocked and treated as malicious bot activity. 
As an alternative, searchsploit~\cite{searchsploit}, the official scraper for ExploitDB's database, was integrated.
While searchsploit is designed for ExploitDB, it traditionally requires manual operation to retrieve or update information. 
Our integration automates these commands, enabling automatic management without human intervention. 
The scraper sends a query to searchsploit, which replies via the standard output. 
The retrieved data is parsed into specific fields: \texttt{exploit name}, \texttt{url}, \texttt{local path}, \texttt{codes}, \texttt{verified}, and \texttt{file type}.\\

For the AlienVault Open Threat Exchange (OTX), the scraper is designed to interact with the API to retrieve "Pulses" associated with available software, operating systems (OSs), and Indicators for CVEs. 
A Pulse provides detailed information about malware, including its description, associated indicators of compromise (IOCs), related pulses, and historical data. 
An Indicator, on the other hand, contains information about potential threats. 
For CVEs, the Indicator includes details such as the CVE description, exploit activity, analysis, related pulses, and user comments. 
API data retrieval is performed using HTTP GET requests, authenticated with an API key. 
However, the number of requests is measured to avoid triggering rate limits, which could result in temporary bans.\\

The OSV (Open Source Vulnerabilities) Database, developed by the Google Security Team, aims to enhance vulnerability triage for open-source software developers.
Its primary objective is to provide precise and actionable data on vulnerabilities, enabling users to identify potential threats and quickly apply security patches.
This project introduced a new schema to improve vulnerability transparency by reducing the maintenance effort for published vulnerabilities and increasing accuracy for downstream consumers.
Unlike other OSINT databases, OSV allows users to query vulnerabilities by specifying a software package or identifier.
This database is initially completely scraped from the updated Google Cloud Storage bucket and subsequently scraped periodically through its API to ensure the most current data is available.\\

After obtaining the requested information from all the previously described sources, the data was preprocessed to eliminate unnecessary details and stored in JSON format in a local PostgreSQL key-value database.
The OSINT scraper was written in Rust due to the security properties guaranteed by the programming language and consists of 1,268 lines of code. 
The scraper is publicly accessible from the group's GitHub repository~\cite{vex_hk}.

\begin{figure}[htbp]
  \centering
  \includegraphics[width=0.8\linewidth]{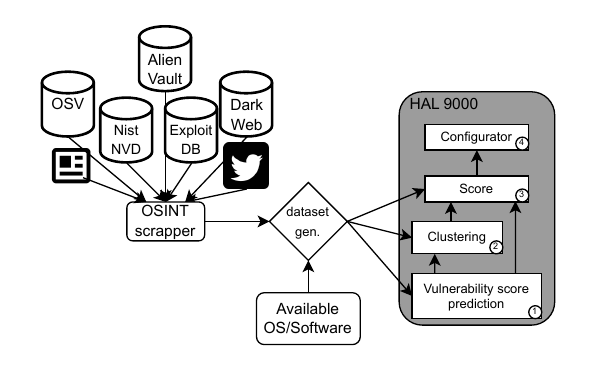}
  \caption{HAL 9000 Architecture and workflow, along with the integration of the OSINT scraper tool. Within HAL's architecture, the data execution path is enumerated from one to four to showcase the execution flow. (The Twitter logo is now officially rebranded as X since 2023.)}
  \label{fig:arch_tool}
\end{figure}

\section{Experiments}\label{sec:performance}
This section outlines the experiments conducted to support the research. 
Initially, three experiments were performed to validate the proposed contributions.
An additional resource analysis was included in this extended work to evaluate the scraper's capabilities and efficiency.

The first experiment aimed to identify the most effective data preprocessing techniques and clustering algorithms for use in a Risk Manager.
In previous work, Lazarus utilized the bag-of-words (BoW) approach for data preprocessing and K-means clustering, employing the elbow method to determine the optimal number of clusters (K).
However, this approach did not provide adequate insights into several key aspects, including the choice of clustering algorithm, dataset size considerations, the selected K value, or strategies for situations where the elbow point is difficult to identify, as shown in Figure~\ref{fig:elbow_method}.
Considering these trade-offs and the advancements discussed in Subsection~\ref{subsec:clustering}, two preprocessing methods—BoW and embeddings—were tested for HAL.
Additionally, twelve different clustering algorithms were applied to the dataset using these preprocessing techniques. 
The resulting clusters were subsequently incorporated into HAL’s implementation to observe their effects on its scoring system.

The second experiment focused on validating HAL's updated scoring system by comparing it with established state-of-the-art methods. 
Specifically, Lazarus's Risk Manager and the approach by Heo et al.~\cite{heo2017designing}, referred to as "Heo" for simplicity. 
The experiment used a dataset containing CVEs recorded up to the end of 2022, simulating a deployment scenario at the beginning of 2023. To mimic real-world conditions, CVEs corresponding to a month of updates were injected into the database during the experiment. 
Sixteen operating systems were evaluated as part of the setup, detailed in the Experimental Setup subsection.

The third experiment evaluated the performance, Root Mean Square Deviation (RMSE), and accuracy of three state-of-the-art methods for CVSS score prediction.
The dataset was used to analyze three aspects of CVSS prediction: metric prediction, severity prediction, and score prediction. 
This experiment sought the most effective approach to predict CVSS scores accurately.

Finally, in the fourth experiment, the performance usage of the scraper was evaluated.
This included measuring the execution time required to retrieve information from various OSINT databases, characterizing the connections to these sources, and identifying additional requirements to ensure successful data retrieval. 
The analysis provided a deeper understanding of the scraper’s efficiency and ability to handle diverse database configurations.


\subsection{Experimental setup}


A dataset containing information on CVEs, vulnerabilities, and exploits related to OSs and software was necessary for the experiments. 
However, no existing tool or dataset provided the required information at the time of writing. 
To address this, a generator was developed to create a dataset with vulnerability information on various OSs and their installed software. 
To ensure a fair comparison with other Risk Managers, the dataset information was limited to entries from the NVD database and ExploitDB, as these were the sources used by the compared Risk Managers.

The dataset included information such as the CVE ID, publication date, last modification date, exploit description, exploit configuration, base metric version 2, and, if available, base metric version 3~\cite{mell2006common}.

Table~\ref{tab:oscve1} and~\ref{tab:oscve2} list the considered OSs available to the Risk Managers from which the dataset was based and the respective amount of considered CVEs.

\begin{table}[htbp]
\centering
\begin{minipage}[t]{0.45\textwidth}
\centering
\caption{List of considered OSs and respective amount of CVEs (Part 1).}
\begin{tabular}{|p{3cm}|p{2cm}|}
\hline
\textbf{OS}                  & \textbf{\#CVEs} \\ \hline
Debian 7            & 3,923  \\ \hline
Windows 10          & 1,514  \\ \hline
FreeBSD 11          & 221    \\ \hline
Debian 8            & 3,923  \\ \hline
Ubuntu 16.04        & 2,035  \\ \hline
Solaris 10          & 359    \\ \hline
OpenBSD 6.0         & 69     \\ \hline
Centos 7            & 8      \\ \hline
Fedora 30           & 835    \\ \hline
Solaris 11          & 359    \\ \hline
Ubuntu 14.04        & 2,035  \\ \hline
Ubuntu 16.04        & 2,035  \\ \hline
\end{tabular}

\label{tab:oscve1}
\end{minipage}
\hfill
\begin{minipage}[t]{0.45\textwidth}
\centering
\caption{List of considered OSs and respective amount of CVEs (Part 2).}
\begin{tabular}{|p{3cm}|p{2cm}|}
\hline
\textbf{OS}                  & \textbf{\#CVEs} \\ \hline
Debian 6            & 3,923  \\ \hline
Ubuntu 17.04        & 2,035  \\ \hline
Fedora 16           & 835    \\ \hline
Ubuntu 12.04        & 2,035  \\ \hline
Ubuntu 22.04        & 2,035  \\ \hline
Debian 10           & 3,923  \\ \hline
Ubuntu 10.04        & 2,035  \\ \hline
Fedora 38           & 835    \\ \hline
Windows Server 2012 & 1,105  \\ \hline
Fedora 24           & 835    \\ \hline
Centos 8            & 8      \\ \hline
OpenSuse 42.1       & 1,298  \\ \hline
\end{tabular}
\label{tab:oscve2}
\end{minipage}
\end{table}




The experiments were conducted on a virtual machine within the XEMU hypervisor, running Debian 12.
The virtual machine had 62 GB of RAM, a 32-core CPU, and an NVIDIA GeForce RTX 3090 graphics card with 24 GB of VRAM.


\subsection{Results and Discussion}

This section presents the results of the experiments and provides a subsequent discussion of the findings. 
The results and respective discussion are organized in the order in which the experiments were conducted.

\begin{figure}[htbp]
  \centering
  \includegraphics[width=\linewidth]{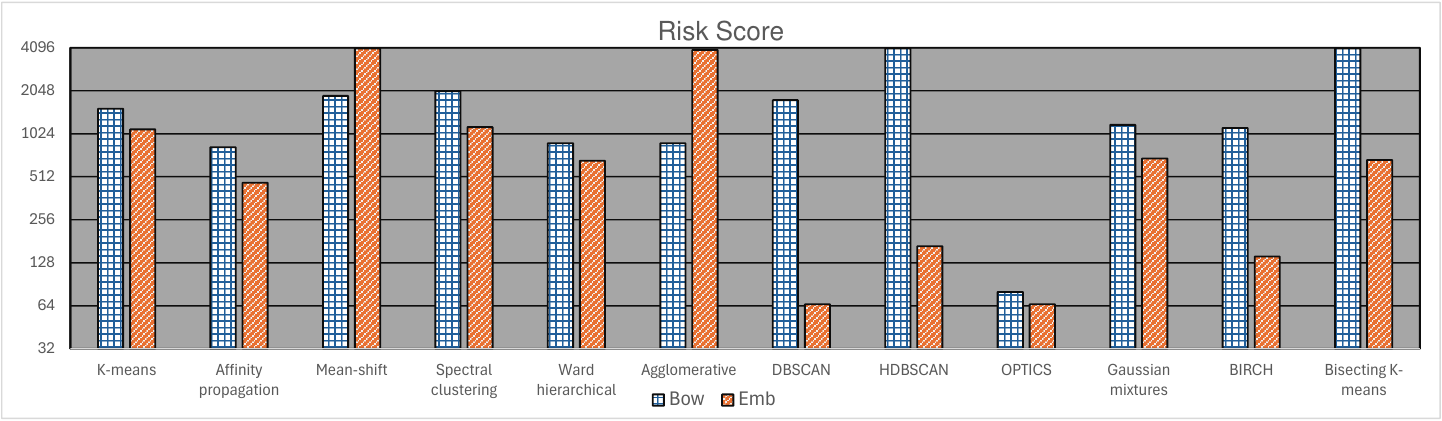}
  \caption{Evaluation of several clustering algorithms and subsequent effects on the HAL risk calculation (lower is better). In each algorithm, two approaches to preprocess data are applied: a bag of words (bow) and sentence embeddings (emb).}
  \label{fig:eval}
\end{figure}

From the experiments conducted with various clustering algorithms, presented in Figure~\ref{fig:eval}, it was observed that sentence embeddings are the optimal preprocessing technique for the given data. 
This technique effectively maintains the relationships between words within sentences.
Regarding the clustering algorithms, it was noticed that OPTICS and DBSCAN provide the lowest risk scores. 
Both algorithms are insensitive to outliers and designed to be scalable, addressing the limitations of the K-means algorithm.
As such, considering these results, HAL utilizes sentence embeddings for preprocessing data and OPTICS for data clusterization.

\begin{figure}[htbp]
  \centering
  \includegraphics[width=\textwidth]{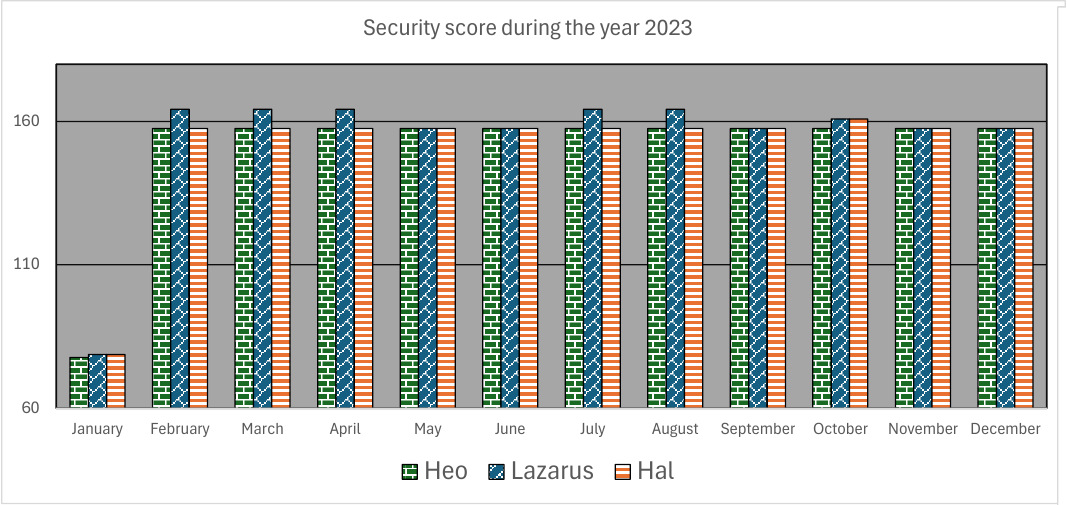}
  \caption{Evaluation of the different Risk Managers, considering the level of security, i.e., the sum of the CVSS score of each CVE present in the advised configuration (lower is better).}
  \label{fig:eval_security}
\end{figure}

\begin{figure}[htbp]
  \centering
  \includegraphics[width=\textwidth]{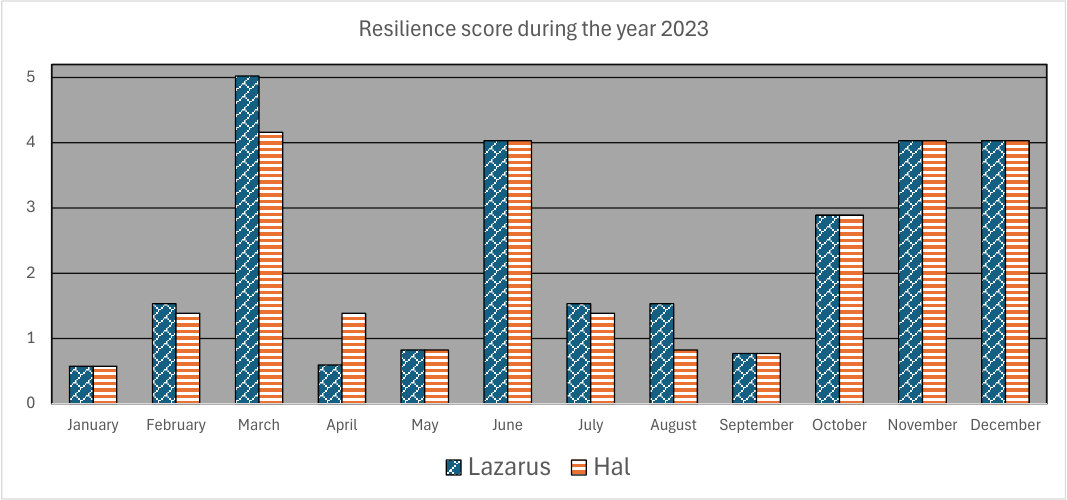}
  \caption{Evaluation of the different Risk Managers, considering the level of resilience, i.e., the multiplication between the reassessed CVSS score of the common CVEs and respective EPSS score (by shared CVE or by clustering) present in the advised configuration (lower is better).}
  \label{fig:eval_resilience}
\end{figure}

We provided two separate graphs for clarity for the risk assessment, presented in Figures~\ref{fig:eval_security} and~\ref{fig:eval_resilience}.
The first graph shows the security score for each Risk Manager throughout 2023, while the second graph displays the resilience score\footnote{The resilience score reflects the number of shared vulnerabilities that can undermine an ITS.
Shared vulnerabilities between nodes increase the threat of parallel attacks.} for each Risk Manager over the same period.
Note that the resilience score for Heo is not calculated in the second graph because it was initially designed for non-replicated nodes. 
Although Heo's implementation was redesigned for the replicated scenario, it does not use a clustering algorithm.
As such, the advised configuration and respective calculation might have hidden shared vulnerabilities that will not show up in its resilience score.
Both graphs use the Lazarus calculation method for the scores presented. 
Each Risk Manager employed the technique to provide the best configuration. 
However, for comparison purposes, we applied the Lazarus scoring method since it is a method that is closer to the other Risk Managers. 
Additionally, we standardized the implementation using the same preprocessing technique and clustering algorithm for the resilience comparison between HAL and Lazarus. 
This comparison aimed to demonstrate that HAL's score calculation system can provide a better configuration than Lazarus.

In the security graph (y-axis), the calculated values represent the sum of the CVSS scores recalculated using Lazarus equations. 
In the resilience graph (y-axis), the values represent the sum of the recalculated CVSS scores multiplied by the sum of the shared CVE EPSS, which indicates the probability of being exploited in the wild. 
This approach was used because a more active vulnerability (with a higher probability of exploitation) can attack several nodes simultaneously, which is not considered in the security calculation.
Initial observations from both graphs indicate a trade-off: Heo provides better security, while Lazarus offers better resilience. However, HAL aims to combine both factors, providing resilient and secure configurations. 
For example, in February, March, July, and August, HAL delivered configurations with lower risk that were resilient and secure compared to Lazarus. 

It is important to note that HAL occasionally produces configurations that are more secure but less resilient than Lazarus, as seen in April. 
This occurred because, in its settings, it was opted to choose more secured configurations over more resilient ones, considering that although the configuration might be exposed to parallel attacks, these would have a lower security score, i.e., have a lesser impact on the node compared with more severe vulnerabilities.

\begin{table}[htbp]
\centering
\caption{Analysis of different state-of-the-art AI implementations for CVSS prediction. Root Mean Square Error (RMSE) is calculated by translating the predicted discrete data into continuous data.}
\resizebox{0.8\textwidth}{!}{%
\begin{tabular}{|l|c|c|c|c|}
\hline
\textbf{Model}          & \textbf{Accuracy (\%)} & \textbf{RMSE} & \textbf{Time Execution (s)} & \textbf{Classifiers}   \\ \hline
Khazaei et al.~\cite{khazaei2016automatic} & 99\%           & 0.66 & 132.85             & Random Forest \\ \hline
Costa et al.~\cite{costa2022predicting}   & 87\%           & 1.55 & 1432.14            & Linear        \\ \hline
VulDistilBERT~\cite{kai2023vuldistilbert}  & 90\%           & 3.63 & 195.49             & Linear       
\\ \hline
\end{tabular}%
}
\label{tab:cvsspred}
\end{table}

For CVSS score prediction, we implemented and tested three methods using the generated dataset from the established state-of-the-art of CVSS prediction.
Each algorithm employs a different technique to predict the CVSS score, as mentioned in the Related Work section.
Although the predicted data is discrete in every implementation, it can be translated into continuous data to calculate the predicted score.
This allows for the computation of the RMSE of the predicted score, which is included in Table~\ref{tab:cvsspred}.

Khazaei et al.\cite{khazaei2016automatic} utilized traditional ML algorithms trained with the descriptions and scores of known CVEs to predict the score. 
Costa et al.\cite{costa2022predicting} applied deep learning algorithms trained with the descriptions and CVSS metrics of known CVEs to predict the CVSS metrics of new vulnerabilities. 
VulDistilBERT~\cite{kai2023vuldistilbert} employed deep learning algorithms trained with CVE descriptions and CVSS severity to predict the CVSS severity of new vulnerabilities.
The results, presented in Table~\ref{tab:cvsspred}, indicate that the method proposed by Khazaei et al. achieves the best accuracy with the provided dataset. 
Additionally, since it uses traditional ML methods (Random Forest), it requires less execution time to train the model and infer new vulnerabilities. 
This implies a quicker response against new vulnerabilities in HAL's use-case scenario.
The method proposed by Costa et al., which predicts each CVSS metric separately, has an average accuracy of around 87\%.
Applying the CVSS v3.1 equations results in calculated CVSS scores with a higher RMSE than the original values. 
It is important to note that this method aimed to aid vulnerability reviewers in speeding up the assessment procedure for new vulnerabilities, which differs from HAL's objective.

Lastly, the method proposed in VulDistilBERT~\cite{kai2023vuldistilbert} achieves 90\% accuracy in predicting new CVEs' severity. 
Although it has high accuracy for severity prediction, mapping it to actual CVSS scores is challenging since severity is relative and spans range: low (0.1 - 3.9), medium (4.0 - 6.9), high (7.0 - 8.9), and critical (9.0 - 10.0). 
This makes predicting precise scores difficult. 
For our calculations, we considered the highest value represented by each severity level to calculate the CVE base score, which explains the high RMSE.
For HAL's implementation, we adopted the method proposed by Khazaei et al.~\cite{khazaei2016automatic} for score prediction.\\


The last experiment evaluated the efficiency of the developed scraper during its initial execution. 
In this first iteration, the scraper retrieves all the available data. 
Subsequent iterations fetch smaller amounts of data, requiring less time and fewer resources.
As a result, the experiment provides an upper bound for the time required to retrieve data.
The results of this experiment are presented in Table~\ref{tab:scraperEval}.

\begin{table}[htbp]
\centering
\caption{Results collected for the scraper retrieving data from NVD, ExploitDB, AlienVault OTX, and OSV. The total execution time includes the data retrieval and insertion into the local database. (* Databases marked with an asterisk are scrapable via file download and upload to the local database.)}
\resizebox{0.8\textwidth}{!}{%
\begin{tabular}{|l|c|c|c|c|c|}
\hline
\textbf{OSINT DB} &
  \textbf{Scrapable} &
  \begin{tabular}[c]{@{}c@{}}\textbf{Total Execution}\\ \textbf{Time (s)}\end{tabular} &
  \begin{tabular}[c]{@{}c@{}}\textbf{API Key}\\ \textbf{Required}\end{tabular} &
  \textbf{\#Entries} &
  \begin{tabular}[c]{@{}c@{}}\textbf{Limitations on}\\ \textbf{\# requests} (req/s)\end{tabular} \\ \hline
NVD            & Yes & 390.29 & Yes & 277,152 & \begin{tabular}[c]{@{}c@{}}50/30 (with key), \\ 5/30 (without key)\end{tabular} \\ \hline
ExploitDB      & Yes*  &    3.42   & No  & 46,575  & N/A                                                                             \\ \hline
AlienVault OTX & Yes & 99,774 & Yes & 277,152     & 10,000/3600                                                                     \\ \hline
OSV            & Yes*  &    235.44    & No  &    255,743     & No limit                                                                            \\ \hline
\end{tabular}%
}

\label{tab:scraperEval}
\end{table}

After the experiment, the total size of the local database was measured, revealing approximately 2 GB of data available for use as a source of local intelligence for HAL.
The most challenging database to retrieve information from was AlienVault OTX due to its hourly request limitation. 
Since the API does not support batch retrieval, requests had to be made individually, which proved time-consuming given the large number of IOCs available on the site. 
This limitation presents a significant challenge when using AlienVault OTX as a source of automated information. 
To address this, a delay of one hour was introduced between requests to ensure fair data retrieval.

Another challenge was the lack of direct data retrieval capabilities from ExploitDB.
To overcome this, we utilized the \textit{searchsploit} tool, which ExploitDB supports for retrieving data.
However, this tool requires periodic updates, which must be synchronized with the scraper to ensure up-to-date information.
Considering the overall execution time (excluding the delays imposed by OTX), the scraper is efficient and can be quickly set up to provide fast vulnerability knowledge.
Its rapid response time allows scalability, making it suitable for more frequent periodic data retrieval schedules.

\section{Conclusion}\label{sec:conclusion}




The extension introduced in this work implements a fully automated scraper that periodically retrieves vulnerability data and cybersecurity intelligence from multiple OSINT sources. 
By broadening the sources of vulnerability information, HAL 9000 significantly enhances the quality and timeliness of input data, leading to more precise and dynamic risk score calculations. 
This development reduces the need for manual intervention, streamlines the workflow, and enables HAL 9000 to make informed assessments in near real-time.

HAL 9000 goes beyond traditional CVE rating systems by leveraging ML techniques to predict CVSS scores for new or unscored vulnerabilities, incorporating factors that traditional methods often overlook, such as patch neglect, exploit likelihood, and the age of CVEs during score reassessment. 
The system’s automated risk scoring and clustering capabilities allow it to distinguish and prioritize vulnerabilities more accurately. 
In fact, experimental results demonstrate that HAL’s clustering of CVEs yields better groupings than prior approaches, facilitating more effective mitigation strategies.

Additionally, HAL 9000 is designed to consider resilience and security risk holistically, allowing it to recommend secure configurations that advance both security posture and operational continuity.
Its independence from a static CVE rating framework enables it to proactively address both known critical vulnerabilities and emerging quasi-zero-day threats, which is particularly valuable in today’s fast-moving threat landscape.

The integration of automated OSINT sources, besides improving vulnerability coverage, also ensures that HAL 9000’s prioritization is continuously informed by the latest available intelligence, enhancing responsiveness to new threats and reducing the window of exposure for unpatched systems. 
This proactive, data-driven approach positions HAL 9000 as a next-generation risk manager, demonstrably superior to traditional systems in terms of automation, accuracy, and adaptability.

HAL 9000’s features, including dynamic CVSS prediction, risk factor incorporation, automated data acquisition, superior CVE clustering, and comprehensive risk assessment, set a new standard for autonomous risk management. 
The demonstrated dependability and extensibility of this approach serve as a roadmap for future systems that harness real-time intelligence and automation to efficiently prioritize and respond to cybersecurity threats.


Future work will focus on further expanding HAL’s functionality by integrating data from penetration testing, automated testing tools, and additional cybersecurity intelligence sources.
The current automated scraper will be enhanced to retrieve insights from additional sources, including the dark web, with retrieved information carefully vetted and incorporated into the local database.

In addition, we will also focus on investigating the potential impact of misinformation on the effectiveness of OSINT integration.
Social media and other open sources may be flooded with fake CVE claims or manipulated by fake accounts and bots that generate a false sense of threat consensus. 
Understanding and addressing these risks will be crucial: future research will examine methods for validating the authenticity and reliability of externally sourced threat data, such as cross-referencing, data provenance tracking, ML–based anomaly detection, or trusted-source weighting, to mitigate the dangers posed by misinformation.

Furthermore, both HAL and the automated scraper will be deployed in real-world ITS implementations~\cite{freitas2023skynet, wang2003sitar, bangalore2009securing, saidane2008design} to assess risk assessment performance and analyze any resulting system overhead. 
These combined efforts aim to further refine HAL’s adaptive security capabilities, ensuring robust, trustworthy, and scalable risk management in evolving cybersecurity environments.

\bmsection*{Author contributions}

Tadeu Freitas led the article's research, implementation, writing, and proofreading. 
Carlos Novo contributed to the writing and proofreading and provided technical advice.
Inês Dutra offered technical advice and proofreading. 
João Soares, Behnam Shariati, and Manuel Correia assisted with proofreading and guidance. 
Rolando Martins supervised the complete work and contributed to the proofreading process.


\bmsection*{Financial disclosure}

The authors Tadeu Freitas and Carlos Novo were supported by the following grants: 2021.04529.BD (FCT) and 2021.08532.BD (FCT), respectively.

\bmsection*{Conflict of interest}

The authors declare no potential conflict of interest.

\bibliography{wileyNJD-AMA}

\end{document}